\documentclass[fleqn,usenatbib,twocolumn,amsmath,prl,superscriptaddress,nofootinbib]{revtex4}
\usepackage{hyperref}
\usepackage{graphicx}
\usepackage{amsmath}
\usepackage{amssymb}
\usepackage{color}

\renewcommand{\vec}[1]{\mathbf{#1}}
\newcommand{\fixme}[1]{{#1}}
\newcommand{\jcap}{JCAP}
\newcommand{\physrep}{Phys.~Rep.}

\begin{document}

\title{\fixme{A first demonstration of CMB delensing using the cosmic infrared background}}

\author{Patricia Larsen}
\email{prl37@cam.ac.uk}
\affiliation{Institute of Astronomy and Kavli Institute for Cosmology Cambridge, Madingley Road, Cambridge, CB3 0HA, UK}
\author{Anthony Challinor}
\affiliation{Institute of Astronomy and Kavli Institute for Cosmology Cambridge, Madingley Road, Cambridge, CB3 0HA, UK}
\affiliation{DAMTP, Centre for Mathematical Sciences, Wilberforce Road, Cambridge CB3 0WA, UK}
\author{Blake D.~Sherwin}
\affiliation{Department of Physics, University of California, Berkeley, CA, 94720, USA}
\affiliation{Miller Institute for Basic Research in Science, University of California, Berkeley, CA, 94720, USA}
\author{Daisy Mak}
\affiliation{Institute of Astronomy and Kavli Institute for Cosmology Cambridge, Madingley Road, Cambridge, CB3 0HA, UK}

\begin{abstract}
Delensing is an increasingly important technique to reverse the gravitational lensing of the cosmic microwave background (CMB) and thus reveal primordial signals the lensing may obscure. We present a first demonstration of delensing on Planck temperature maps using the cosmic infrared background (CIB). Reversing the lensing deflections in Planck CMB temperature maps using a linear combination of the 545 and 857\,GHz maps as a lensing tracer, we find that the lensing effects in the temperature power spectrum are reduced in a manner consistent with theoretical expectations. In particular, the characteristic sharpening of the acoustic peaks of the temperature power spectrum resulting from successful delensing is detected at a significance of $16\,\sigma$, with an amplitude of $A_{\mathrm{delens}}=1.12\pm 0.07$ relative to the expected value of unity. This first demonstration on data of CIB delensing, and of delensing techniques in general, is significant because lensing removal will soon be essential for achieving high-precision constraints on inflationary B-mode polarization.
\end{abstract}

\maketitle
\section{INTRODUCTION}

The gravitational lensing of the cosmic microwave background (CMB) is rapidly becoming a powerful probe of cosmology and fundamental physics \citep{smith07,act1, spt1,plancklens,polarbear,biceplens}. However, the lensing deflections are not only a source of information, but they can also obscure signals in the primordial CMB. In particular, gravitational lensing converts E-mode into B-mode polarization; these lensing-induced B-modes act as a source of noise~\cite{Lewis2002,Hu2002} that limits constraints on the primordial B-mode power from inflationary gravitational waves \citep{KKS97,ZS98}. Gravitational lensing also smooths the acoustic peaks in the temperature power spectrum, which weakens constraints on cosmological parameters probed by the peak positions, such as the number of relativistic particle species. With the rapid increase in sensitivity expected from forthcoming CMB experiments, even near-future CMB B-mode searches will be completely lensing-limited. Delensing, the process of reversing, or removing, the lensing effects in maps, is therefore a method that will be essential to realizing powerful constraints on inflationary B-modes and other parameters~\cite{KnoxSong2002,KCK2002,seljakhirataBmodes}. However, despite their importance for progress in key areas of CMB research and despite much theoretical work, delensing methods have thus far not been successfully demonstrated on data. In this Letter, we show a first demonstration of delensing with CMB temperature data from Planck.

The method generally considered to delens is to take a field that traces the dark matter distribution and filter it to obtain a weighted proxy for the true lensing map. This is then used to reverse the lensing in CMB maps, or, equivalently in the case of B-mode delensing, construct a linearized estimate of the lens-induced modes that is subtracted off (see, e.g., Refs.~\cite{HuOkamoto2002,KnoxSong2002,KCK2002,seljakhirataBmodes,Smith2009,delensingexternal,delensinginflation,SKAdelensing}). Though often CMB-internal delensing (i.e., using a lensing map reconstructed from the CMB itself) is considered, Ref.~\cite{sherwin15} discuss in detail the use of the cosmic infrared background (CIB) for this purpose; see also~\cite{delensinginflation}. The CIB is mostly made up of unresolved emission from dusty star-forming galaxies at high redshifts; as the CMB lensing arises from similar high redshifts, the CIB should be strongly correlated with the lensing convergence. This correlation has been measured to be up to around $80\%$, suggesting that one could remove more than half of the lensing~\citep{sherwin15}, assuming the CIB can be accurately separated from Galactic foregrounds.

In this paper, we use the CIB estimated from Planck maps at 545 and 857\,GHz
to delens the CMB temperature anisotropies measured by Planck at lower frequencies.
In particular, we measure the CMB temperature angular power spectrum after delensing and show that the lensing-induced peak-smoothing is significantly reduced in agreement with expectations. Aside from being of general interest as a first demonstration of delensing methods applied to data, this is useful for additional reasons. First, it provides confirmation that CIB delensing is possible despite challenges such as accurately separating the CIB and Galactic dust emission.
Second, our demonstration that the change in the power spectrum by delensing is consistent with expectations is a model-independent test that lensing affects the temperature power spectrum in the expected manner. 

\section{THEORY}
\label{sec:theory}

In the flat-sky approximation, gravitational lensing alters the CMB temperature map by the transformation
\begin{equation}
T(\vec{x}) = \tilde{T}(\vec{x}+\boldsymbol{\nabla} \phi ) \, , 
\label{eq:lens1}
\end{equation}
where $\phi$ is the CMB lensing potential evaluated at position $\vec{x}$,
and $\tilde{T}$ is the unlensed temperature. The effect
 of this transformation on the temperature power spectrum is a smoothing of the acoustic peaks and a transfer of power to small scales.
See, e.g., Ref.~\cite{lewis06} for a detailed review of CMB lensing. 

Delensing attempts to undo the remapping in Eq.~\eqref{eq:lens1} by applying the reverse transformation
\begin{equation} 
\tilde{T}(\vec{x}) = T(\vec{x} - \tilde{\boldsymbol{\alpha}}) \, .
\label{eq:transform}
\end{equation}
Here, $\tilde{\boldsymbol{\alpha}}$ should correspond to the gradient of the lensing potential at the position $\tilde{\vec{x}}$, where $\tilde{\vec{x}} + \tilde{\boldsymbol{\alpha}} = \vec{x}$. In practice, we find that using $\tilde{\boldsymbol{\alpha}} \approx \left. \boldsymbol{\nabla}\phi \right|_{\vec{x}}$ is a very good approximation since the lensing deflections relevant for smoothing the temperature power spectrum are a few arcminutes, but are coherent on degree scales.

Ideally, one would construct $\tilde{\boldsymbol{\alpha}}$ from a reconstruction of the lensing potential from the CMB itself (e.g., Ref.~\cite{HuOkamoto2002}). In practice, current lens reconstructions are rather noisy so instead we use a filtered version of the CIB as a proxy for the lensing potential. Although the CIB is not fully correlated with the lensing convergence, it can be measured with high signal-to-noise on degree scales. Denoting the CIB map as $I(\vec{x})$, we construct an estimate of the lensing potential by applying the Wiener filter $F^I_\ell = C_\ell^{I\phi}/C_\ell^{II,\text{tot}}$ in harmonic space at multipole $\ell$. Here, $C_\ell^{I\phi}$ is the CIB-lensing cross-correlation and $C_\ell^{II,\text{tot}}$ is the total power spectrum of $I(\vec{x})$, which includes clustering of the CIB, shot noise, instrumental noise and any other contaminants such as residual Galactic emission.

Denoting the Wiener-filtered CIB by $I^F(\vec{x})$, we construct the delensed CMB map as
\begin{equation}
T^{\text{delens}}(\vec{x}) = T(\vec{x} - \boldsymbol{\nabla} I^F) \, .
\label{eq:delens}
\end{equation}
This is approximately the unlensed CMB at the position $\vec{x} + \boldsymbol{\nabla}(\phi - I^F)$, so the delensed map contains only the residual lensing from $\boldsymbol{\nabla}(\phi - I^F)$. Generally, the power spectrum of the lensing potential is effectively reduced by delensing as
\begin{equation}
C_\ell^{\phi\phi} \rightarrow C_\ell^{\phi\phi} - 2 F_\ell^I C_\ell^{I\phi} + \left(F_\ell^I\right)^2 C_\ell^{II,\text{tot}} \, .
\label{eq:residpower}
\end{equation}
This residual power is minimised if the filter $F^I_\ell$ is chosen as above to be the Wiener filter. In this case, the residual power is $C_\ell^{\phi\phi}\left(1-\rho_\ell^2\right)$, where the correlation coefficient
\begin{equation}
\rho_\ell = \frac{C_\ell^{I\phi}}{\sqrt{C_\ell^{II,\text{tot}} C_\ell^{\phi\phi}}} \, .
\label{eq:residpowerWiener}
\end{equation}
In practice, there will be some mismatch between the spectra used in the filter and the true spectra that appear in Eq.~\eqref{eq:residpower}. As it can be shown that the mismatch only affects the residual lensing power at second order, we shall ignore this small source of error.

We make use of the lensing implementation in the \textsc{CAMB} Boltzmann code~\cite{CAMB}, with the replacement $C_\ell^{\phi\phi} \rightarrow C_\ell^{\phi\phi}\left(1-\rho_\ell^2\right)$, to calculate the expected change in the temperature 
power spectrum, using the parameters of the best-fit Planck 2015 $\Lambda$CDM cosmology~\cite{2015arXiv150201589P} to compute the spectra of the unlensed CMB and the lensing potential. Our modelling of $\rho_\ell$ is discussed below.

\section{DATA}
\label{sec:data}

We reconstruct the CIB from the Planck 545 and 857\,GHz full-mission maps. For the CMB, we use the \textsc{SMICA} reconstructions that linearly combine the Planck frequency maps to reduce foreground contamination~\cite{2015arXiv150205956P}. 
To check our modelling of $\rho_\ell$, we use the Planck 2015 lensing reconstruction~\cite{2015arXiv150201591P}. We make use of the Planck full focal-plane simulations~\cite{2015arXiv150906348P} to validate our methodology and to estimate our final error bars.

\section{METHODOLOGY}
\label{sec:methodology}

\begin{figure}
\includegraphics[width=0.5\textwidth]{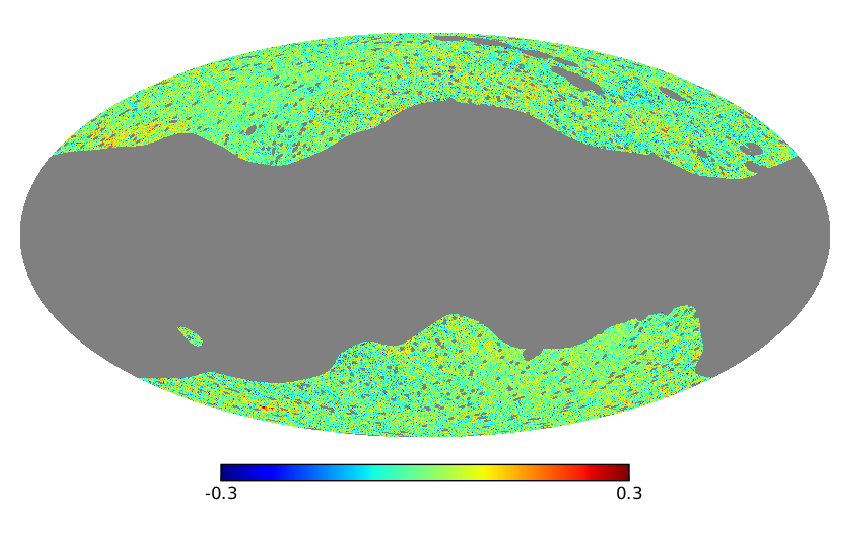}
\caption{Map of the CIB anisotropies from the frequency combination $77 M^{545}- M^{857}$  in units of Kelvin. The mask used in our analysis, before apodization, is shown in grey. }
\label{fig:map}
\end{figure}

The 545 and 857\,GHz Planck maps are dominated by Galactic dust, particularly on the degree scales that are relevant for delensing the CMB temperature. However, the dust emission is highly correlated between these frequencies so we can remove the majority of it\footnote{\fixme{%
For the masks that we use here, at multipoles $\ell \approx 100$ the ratio of (residual) Galactic dust power to clustered CIB in $I(\vec{x})$ is around 1:2, compared to 15:1 in the original 545\,GHz map.}} by taking the frequency combination (in CMB temperature units)
$I(\vec{x}) = 77 M^{545}(\vec{x}) - M^{857}(\vec{x})$. Here, $M^\nu(\vec{x})$ is the Planck full-mission map at frequency $\nu$, and the coefficient in the linear combination follows from a likelihood analysis of the \fixme{auto- and cross-frequency spectra of the Planck high-frequency maps~\cite{daisy}, which fits component models to each spectrum}. The map $I(\vec{x})$ is shown in Fig.~\ref{fig:map}.

\begin{figure}
\includegraphics[width=0.5\textwidth]{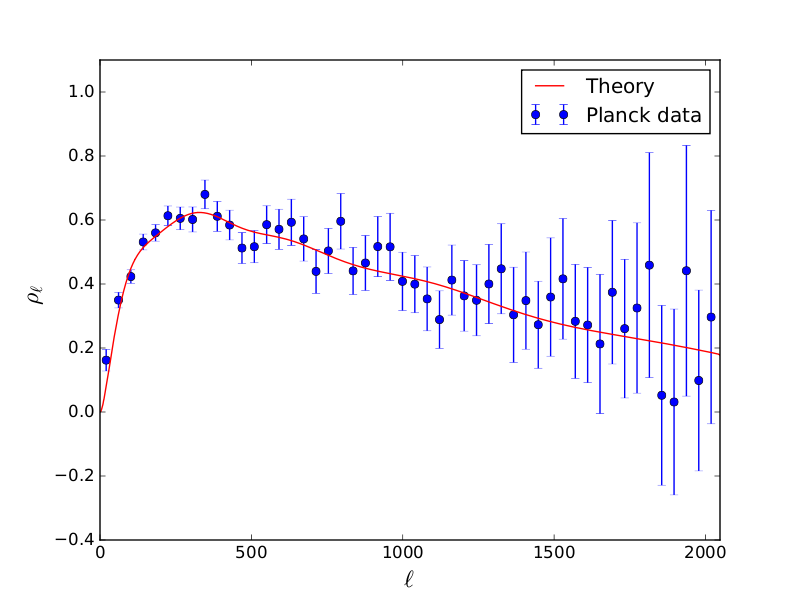}
\caption{Correlation coefficient measured between our CIB map $I(\vec{x})$ and the Planck 2015 lensing potential reconstruction. The correlation is normalised by the theory $C_\ell^{\phi\phi}$. The correlation is measured in bins of width $\Delta \ell = 40$,
and the error bars are estimated from the scatter within the bins. We use a model for $\rho_\ell$, shown as the red line, when constructing the expected power after delensing and for the Wiener filter $F^I_\ell$.}
\label{fig:coeff}
\end{figure}

For the cross-correlation $C_\ell^{I\phi}$, we use the model from Ref.~\cite{2014A&A...571A..18P}, which is fit there to the 2013 Planck data and which we recalibrate to account for calibration changes from the 2013 to 2015 maps. We use a multi-component model for $C_\ell^{II,\text{tot}}$, consisting of clustered CIB, shot noise, instrument noise and Galactic dust residuals, that we fit to the measured auto-power spectrum of $I(\vec{x})$ taking account of our masking procedure described below. We combine these models with the power spectrum of the lensing potential to obtain the correlation coefficient $\rho_\ell$. This model for $\rho_\ell$ is compared to the correlation that we measure between $I(\vec{x})$ and the Planck 2015 lensing reconstruction in Fig.~\ref{fig:coeff}. In our analysis, we discard multipoles of the CIB with $\ell < 50$ to reduce the impact of inaccuracies in our modelling of the (non-Gaussian) dust residuals. However, we note that extending this cut to lower multipoles in future analyses may allow more of the lensing to be removed.

\fixme{We carry out our delensing procedure using the interpolation scheme of the publicly-available Lenspix code~\cite{2005PhRvD..71h3008L}. This procedure
uses the gradient of the filtered CIB map, $I^F(\vec{x})$, to remap separately the \textsc{SMICA} CMB maps from the first and second half of the mission following Eq.~\eqref{eq:delens}. 
These are then cross-correlated using the PolSpice code~\cite{2004MNRAS.350..914C} to obtain an estimate of the delensed power spectrum that is free of noise bias.}

Initially a mask comprising the union of the Planck 40\,\% Galactic mask, the Planck point source mask made up of sources detected with $S/N > 5$ at 100\,GHz in the 
PCCS2 and PCCS2E catalogues~\cite{2015arXiv150702058P}, and an
additional mask removing remaining areas of visible dust contamination in the frequency-differenced map $I(\vec{x})$ that arise from local strong variations in the dust spectral energy distribution, is applied to the CIB map. This mask is shown in grey in Fig.~\ref{fig:map}. The Galactic mask is apodized by iteratively smoothing with a 60-arcmin.\ Gaussian, 
while the other mask components use a simple cosine apodization. 
To prevent these apodized edges from impacting the gradient of the filtered CIB map, we apply a further binary mask removing regions close to the apodized edges after computing the gradient. 
Finally, we apply a similar apodization procedure to this extended mask and use this to mask the CMB temperature maps after delensing. We have confirmed that adding a mask that is the union of point sources detected at 143\,GHz and 217\,GHz to the CMB temperature maps
has no significant impact on our results.  

\subsection{Simulations}

\begin{figure}
\includegraphics[width=0.5\textwidth]{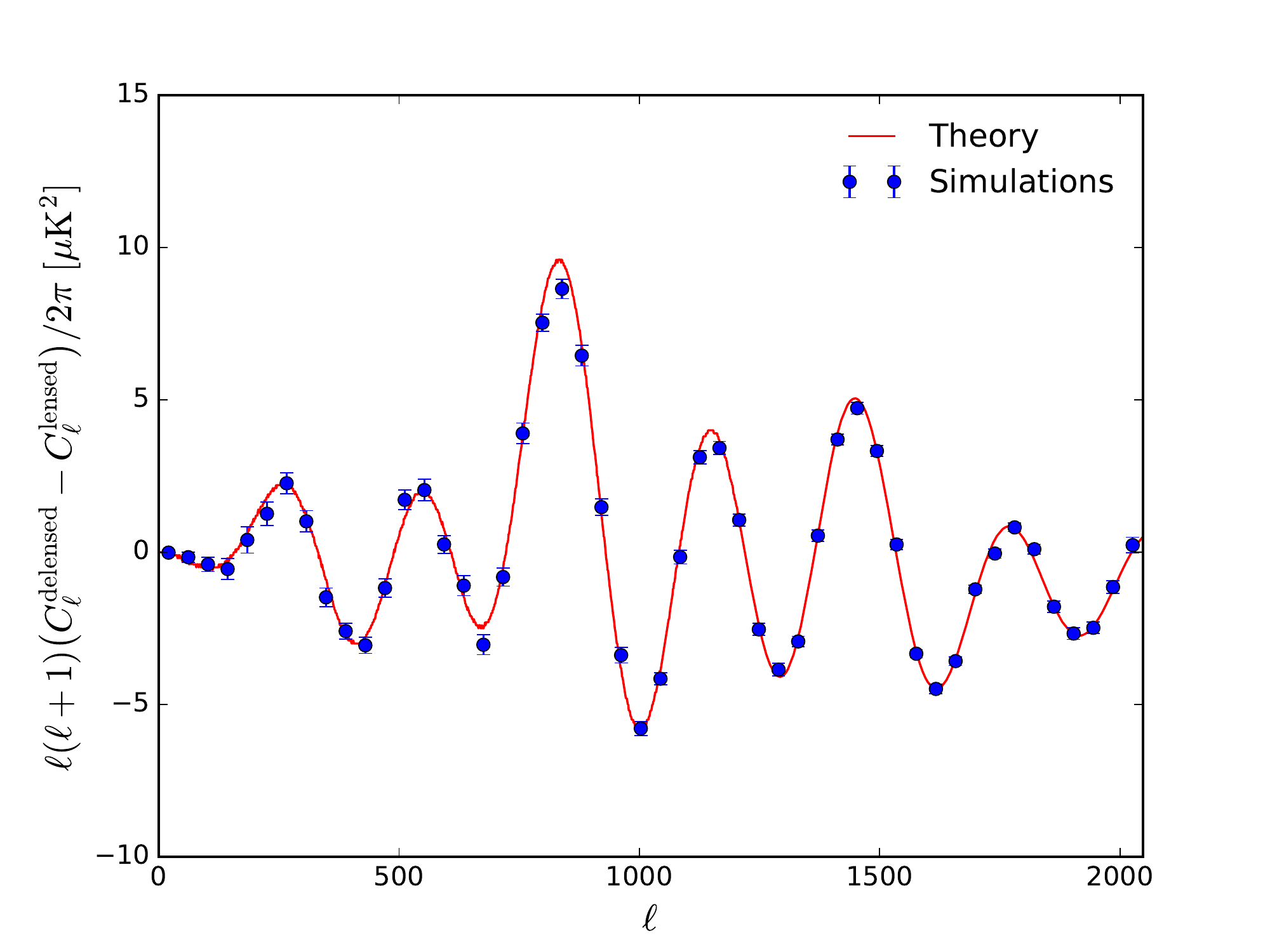}
\caption{Mean difference between the delensed and original CMB spectra for the 100 simulations described in the text. The error bars are the standard error on the mean and are estimated from the scatter across the simulations. The expected difference in power is shown as the red line.} 
\label{fig:sims}
\end{figure}

We test this delensing procedure using 100 Gaussian CIB simulations that are appropriately correlated with the lensing potential fields used in the Planck CMB simulations. The CIB simulations are used to delens 100 pairs of \textsc{SMICA} half-mission CMB maps from the Planck full focal-plane simulation suite using the same filters and masks that we use to delens the data. The mean difference between the power spectra of the delensed maps and the original maps (with the same mask) is plotted against the theoretical expectation in Fig.~\ref{fig:sims}. The simulation results are in good agreement with the expectation, with any bias due to, e.g., interpolation errors well below the statistical error on a single realisation.

\section{RESULTS}
\label{sec:results}

\begin{figure}
\includegraphics[width=0.5\textwidth]{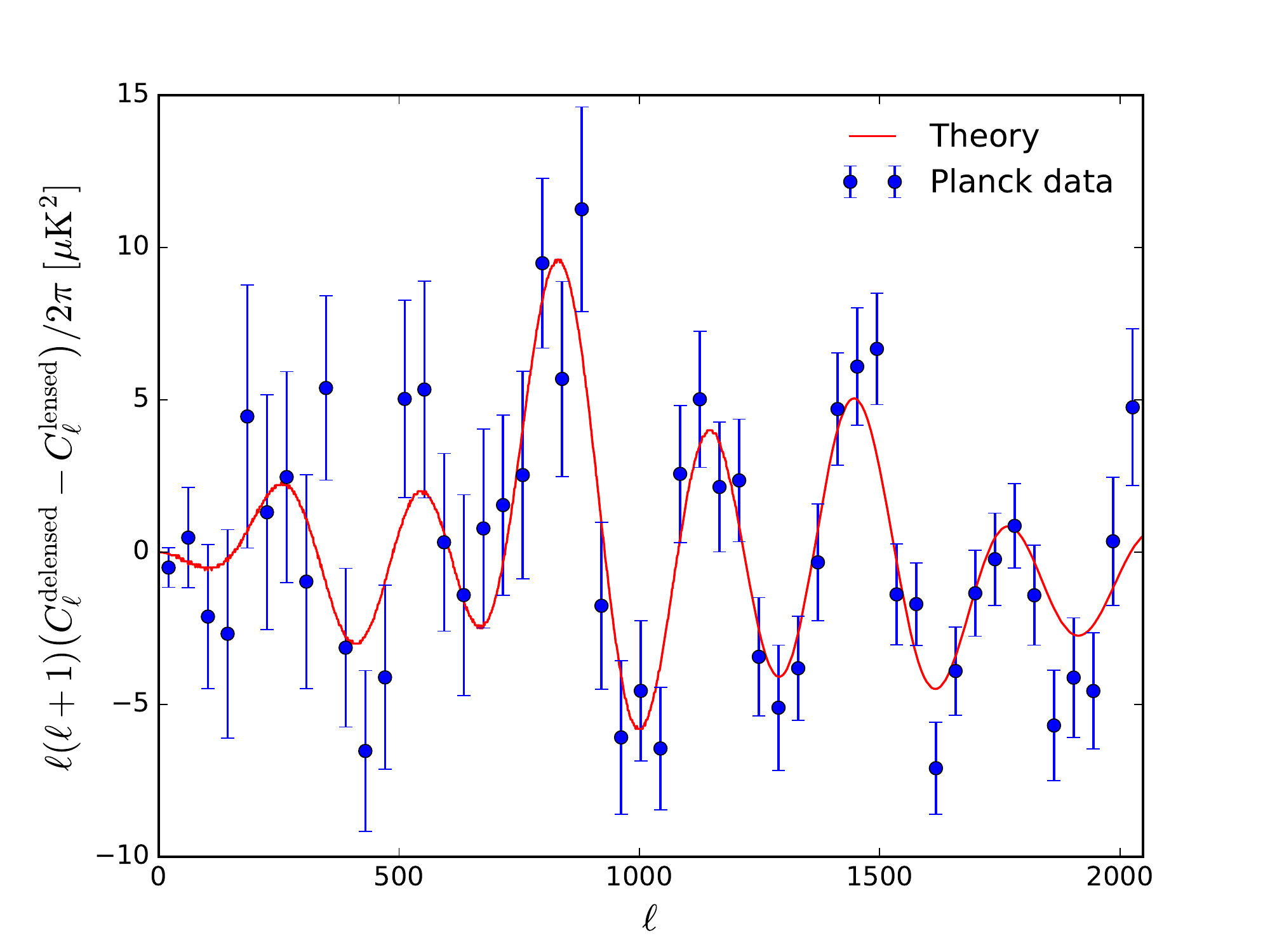}
\caption{Difference between the power spectra of the CIB-delensed and original CMB maps, binned with $\Delta \ell = 40$. The error bars are computed from the scatter across simulations. The expected power difference is shown as the red line.}
\label{fig:results}
\end{figure}

The difference we measure between the delensed CMB power spectrum and the power spectrum before delensing is shown in Fig.~\ref{fig:results}. The error bars are calculated using the simulations described above. We test the significance of our measurement by fitting for an amplitude parameter, $A_{\mathrm{delens}}$, which scales the expected difference in power (i.e., the red line in Fig.~\ref{fig:results}).
We do this with a simple $\chi^2$ minimization using the full covariance matrix and obtain a value of $A_{\mathrm{delens}} = 1.12\pm 0.07$, consistent with the expected value of unity. We detect the change in power due to delensing at approximately $16\,\sigma$. The reduced $\chi^2/\text{d.o.f.} = 1.35$ for 49 degrees of freedom at the best-fit $A_{\text{delens}}$ model, giving a marginally acceptable fit (a probability-to-exceed of 5\%). This somewhat high $\chi^2$ value appears to be due primarily to covariances involving the single point around $\ell\approx 350$. We note further that our CIB simulations model any astrophysical component uncorrelated with $\phi$ as a statistically-isotropic, Gaussian random field, which is incorrect in detail for dust residuals. This may have a small impact on both the expected value of the delensed power spectrum and the error bars.
Finally, by fitting the power spectrum differences in Fig.~\ref{fig:results} to a 
template model proportional to the difference between the \emph{unlensed} and lensed spectra, we find that delensing removes approximately 20\,\% of the lensing power, consistent with the typical values of $\rho_\ell^2$ on large scales (see Fig.~\ref{fig:coeff}).\footnote{\fixme{We note that we detect the effects of delensing here at higher significance than lensing itself is detected in the temperature power spectrum (via the phenomenological $A_{\text{L}}$ parameter~\cite{2015arXiv150201589P}). This is despite the difference between the lensed and unlensed spectra being around five times larger than between the lensed and CIB delensed spectra. The higher significance here is due to a cancellation of cosmic variance when taking the difference between the lensed and delensed spectra. (Degeneracies between $A_{\text{L}}$ and other cosmological parameters further limit the precision of estimates of $A_{\text{L}}$.)}}
We note that the reduction in lensing power is multipole dependent; 20\,\% is an average value for CMB multipoles $\ell \leq 2048$. 

\begin{figure}
\includegraphics[width=0.5\textwidth]{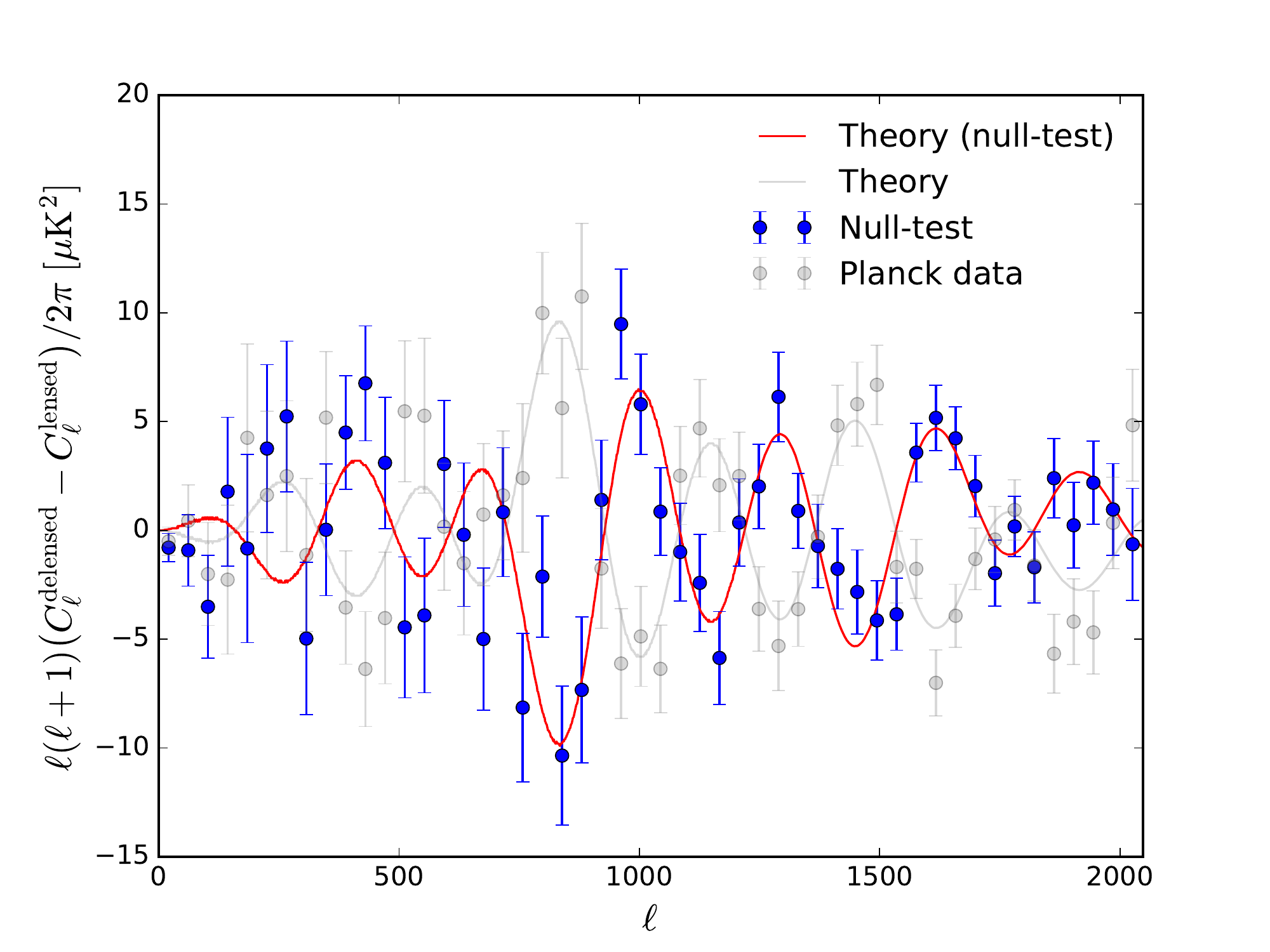}
\caption{As Fig.~\ref{fig:results}, but for the null-test in which the CMB map is flipped in the plane of the Galaxy prior to CIB delensing. This procedure adds uncorrelated lenses giving the expected change in the CMB power spectrum shown as the red line.
Our main delensing result is shown in the background to highlight the opposite sign of the change in the CMB power spectrum for the null-test.}
\label{fig:nulltest}
\end{figure}

In Fig.~\ref{fig:nulltest} we present the results of a null-test whereby we flip the
CMB temperature map over the plane of the Galaxy, but not the CIB map, before delensing. The CIB is uncorrelated with the lensing potential in the flipped CMB map so ``delensing'' adds lensing power rather than subtracting it: $C_\ell^{\phi\phi} \rightarrow C_\ell^{\phi\phi} + (F_\ell^I)^2 C_\ell^{II,\text{tot}} = C_\ell^{\phi\phi}(1+\rho_\ell^2)$. We use the same masking procedure as for the main analysis since the mask should still be 
adequate to remove any significant Galactic foreground residuals in the flipped CMB map. Fitting an amplitude $A_{\text{null}}$ that scales the expected change in the CMB power spectrum for this null-test, we find $A_{\rm{null}} = 0.86 \pm 0.07$, 
where $A_{\rm{null}}=1$ is the expected value. \fixme{This confirms that the observed reduction in peak smoothing from our delensing procedure is due to the correlations between the observed CIB and the particular realisation of lenses that are imprinted in the observed CMB.}

\section{SUMMARY AND CONCLUSIONS}
\label{sec:summary}

We have presented a first demonstration of CIB delensing on the temperature anisotropies of the CMB. We demonstrate the expected sharpening of the acoustic peaks at high significance, reporting a $16\,\sigma$ detection of delensing effects in the power spectrum. We note that this sharpening of the peaks is a robust effect that is not easily mimicked by instrumental or astrophysical systematics, which
would generally add uncorrelated lensing effects. This is illustrated by the fact that null-tests involving delensing with uncorrelated fields result in a smoothing rather than sharpening of the peaks (i.e., lensing-like rather than delensing effects).

The reduction in the peak-smoothing of the temperature spectrum is found to be in agreement with the theoretical expectation (based on the correlation coefficient of our CIB maps with the Planck reconstruction of the lensing potential).  In particular, the delensing strength is found to be $A_{\mathrm{lens}}= 1.12 \pm 0.07$ relative to a fiducial expectation of unity. This agreement of data and theory shows consistency between the CMB power spectrum analysis and CIB-derived lensing potential maps.  We note that, beyond a consistency test, CIB delensing of the temperature and, particularly, E-mode polarization may allow for improved constraints on parameters such as the effective number of relativistic degrees of freedom ($N_{\text{eff}}$),
but we defer such considerations to future work.
 
More broadly, our results demonstrate the viability of CIB delensing as a method. However, while our simple multi-frequency method for Galactic dust removal is sufficient for our current purposes, it results in a correlation coefficient of the CIB with the lensing potential of only around 40\,\% on the degree scales relevant for delensing the temperature anisotropies. The correlation is stronger on the (smaller) scales that are relevant for delensing B-modes, but ultimately more sophisticated methods, e.g., the approach of Ref.~\cite{2016arXiv160509387P} or $\text{H}_{\text{\textsc{I}}}$-based dust cleaning, will be needed
to achieve delensing performance closer to that forecasted in Ref.~\cite{sherwin15}.
 
With upcoming CMB experiments, delensing methods will be crucial for revealing small inflationary B-mode polarization signals; our work on CIB-delensing of the CMB temperature anisotropies is an early step in this important research area.

\begin{acknowledgments}
PL is supported jointly by the Royal Society of New Zealand Rutherford Foundation Trust and the Cambridge Commonwealth Trust. 
We thank Mark Ashdown and Julian Borrill for organising access to the Planck FFP8 simulations, and Jo Dunkley, Antony Lewis, Neelima Sehgal and Alexander van Engelen
for useful conversations. Some of the results in this paper have been derived using the HEALPix \cite{healpix} package.
\end{acknowledgments}

\bibliographystyle{apsrev4-1}
\bibliography{larsen_updated,challinor}
\end{document}